\documentclass{nature}

\usepackage{lettrine}
\usepackage{graphicx}
\usepackage{dcolumn}
\usepackage{bm}
\usepackage{amssymb}
\usepackage{amssymb}
\usepackage{epsfig}
\usepackage[10pt]{moresize}
\usepackage{comment}
\usepackage{graphicx}
\usepackage{epsfig}
\usepackage{subfigure}

\title{Multi-target-qubit unconventional geometric phase gate in a multi-cavity system}
\author{Tong Liu, Xiao-Zhi Cao, Qi-Ping Su, Shao-Jie Xiong $\&$  Chui-Ping Yang$^{\star}$}
\begin{document}

\maketitle
\begin{affiliations}
\item[] Department of Physics, Hangzhou Normal University,
Hangzhou, Zhejiang 310036, China \\
$^\star$e-mail:yangcp@hznu.edu.cn
\end{affiliations}

\begin{abstract}
Cavity-based large scale quantum information processing (QIP) may involve multiple cavities and require performing various quantum logic operations on qubits distributed in different cavities. Geometric-phase-based quantum computing has drawn much attention recently, which offers advantages against inaccuracies and local fluctuations. In addition, multiqubit gates are particularly appealing and play important roles in QIP.  We here present a simple and efficient scheme for realizing a multi-target-qubit unconventional geometric phase gate in a multi-cavity system. This multiqubit phase gate has a common control qubit but different target qubits distributed in different cavities, which can be achieved using a single-step operation. The gate operation time is independent of the number of qubits and only two levels for each qubit are needed. This multiqubit gate is generic, e.g., by performing single-qubit operations, it can be converted into two types of significant multi-target-qubit phase gates useful in QIP and quantum Fourier transform. The proposal is quite general, which can be used to accomplish the same task for a general type of qubits such as atoms, NV centers, quantum dots, and qubits.
\end{abstract}


\lettrine[lines=2]{M}ultiqubit gates are particularly appealing and have
been considered as an attractive building block for quantum information
processing (QIP). There are two types of significant multiqubit gates, i.e.,
multiqubit gates with multiple control qubits acting on a single target
qubit~[1-6], and multiqubit gates with a single qubit simultaneously
controlling multiple target qubits~[7]. These two kinds of multiqubit gate
play important roles in QIP. For instance, a multi-control-qubit gate has
applications in quantum algorithms~[8,9], error correction~[10,11], and
quantum Fourier transform~[12]; and a multi-target-qubit gate is useful in
quantum cloning~[13], quantum algorithms~[8,12,14], entanglement
preparation~[15], and error correction~[16].

A multiqubit gate can in principle be constructed by using single-qubit and
two-qubit basic gates. However, when using the conventional
gate-decomposition protocols to construct a multiqubit gate [17,18], the
number of basic gates increases and the procedure usually becomes
complicated as the number of qubits increases. Hence, building a multiqubit
gate may become very difficult since each basic gate requires turning on and
off a given Hamiltonian for a certain period of time, and each additional
basic gate adds experimental complications and the possibility of more
errors. During the past years, there is much interest in \textit{directly}
implementing multiqubit gates. Proposals have been presented for directly
realizing both multi-control-qubit gates [1-6] and multi-target-qubit gates
[7] in various physical systems. Note that the gate implementation using
these previous proposals [1-7] was based on non-geometric dynamical
evolution.

In recent years, much attention has been paid to fault tolerant quantum
computing, which operates essentially based on geometric phase~[19-22]. The
geometric phase is determined by the global features of the evolution path,
which offers potential advantages against local fluctuations. A great deal
of work has been devoted to construct geometric phase gates, which are
usually divided into two categories: conventional geometric phase gates and
unconventional geometric phase gates. The former needs to remove the
dynamical phase by choosing the cyclic evolution in dark states~[23] or
employing multi-loop schemes (the evolution is driven by a Hamiltonian along
several closed loops)~[24,25]. But the latter does not require performing
\textit{additional} operations to cancel the dynamical phase, because the
total phase is dependent only on global geometric features and independent
of initial states of the system~[26,27]. The unconventional geometric phase
gate was first proposed in~[26]. According to~[26], an unconventional
geometric phase gate is characterized by a unitary operator $U$($\{\gamma \}$%
), where $\gamma $ is the total phase, which consists of a geometric phase
and a dynamic phase (see [26]).

During the past years, proposals for realizing both conventional geometric
phase gates [23-25,28-31] and unconventional geometric phase gates
[26,27,32] were presented. Moreover, high fidelity conventional geometric
phase gates have been experimentally demonstrated using NMR~[33],
superconducting qubits~[34], and nitrogen-vacancy (NV) centers~[35,36]. In
addition, a two-qubit unconventional geometric phase gate has been
experimentally realized in a trapped ion system~[37]. However, previous
works are mainly focused on how to construct single-qubit or two-qubit
conventional/unconventional geometric phase gates~[23-37], or implementing a
multi-control-qubit gate~[1-6] and a multi-target-qubit gate~[7] based on
non-geometric dynamical evolution.

In this work, we consider how to implement a multi-target-qubit
unconventional geometric phase gate, which is described by the following
transformation:
\begin{eqnarray}
|+\rangle _{A}\prod\limits_{j=1}^{n}|i_{j}\rangle &\rightarrow &|+\rangle
_{A}\prod\limits_{j=1}^{n}e^{i\theta _{j}\left\langle +\right\vert \left.
i_{j}\right\rangle }|i_{j}\rangle ,  \nonumber \\
|-\rangle _{A}\prod\limits_{j=1}^{n}|i_{j}\rangle &\rightarrow &|-\rangle
_{A}\prod\limits_{j=1}^{n}e^{i\theta _{j}\left\langle -\right\vert \left.
i_{j}\right\rangle }|i_{j}\rangle ,
\end{eqnarray}%
where subscript $A$ represents a control qubit, subscripts ($1,2,...,n$)
represent $n$ target qubits ($1,2,...,n$), and $\prod\limits_{j=1}^{n}|i_{j}%
\rangle $ (with $i_{j}\in \left\{ +,-\right\} $) is the $n$-target-qubit
computational basis state. For $n$ target qubits, there are a total number
of $2^{n}$ computational basis states, which form a set of complete
orthogonal bases in a $2^{n}$-dimensional Hilbert space of the $n$ qubits.
Equation~(1) shows that when the control qubit $A$ is in the state $%
\left\vert +\right\rangle $ ($\left\vert -\right\rangle $), a phase shift $%
e^{i\theta _{j}}$ happens to the state $\left\vert +\right\rangle $ ($%
\left\vert -\right\rangle $) but nothing happens to the state $\left\vert
-\right\rangle $ ($\left\vert +\right\rangle $) of the target qubit $j$ ($%
j=1,2,...,n$). For instance, under the transformation~(1), one has: (i) the
state transformation described by following Eq.~(18) for a two-qubit phase
gate on control qubit $A$ and target qubit $j$, and (ii) the state
transformation described by Eq.~(21) below for a three-qubit phase gate on
control qubit $A$ and two target qubits ($1,2$). Note that the multiqubit
phase gate described by Eq.~(1) is equivalent to such $n$ two-qubit phase
gates, i.e., each of them has a common control qubit $A$ but a different
target qubit $1,2,...,$ or $n,$ and the two-qubit phase gate acting on the
control qubit $A$ and the target qubit $j$ ($j=1,2,...,n$) is described by
Eq.~(18).

The multiqubit gate described by Eq. (1) is generic. For example, by
performing a single-qubit operation such that $|+\rangle _{A}\rightarrow
\prod\limits_{j=1}^{n}e^{-i\theta _{j}}|+\rangle _{A}$ and $|-\rangle
_{j}\rightarrow e^{i\theta _{j}}|-\rangle _{j}$ but nothing to $|-\rangle
_{A}$ and $|+\rangle _{j},$ the transformation~(1) becomes
\begin{eqnarray}
|+\rangle _{A}\prod\limits_{j=1}^{n}|i_{j}\rangle  &\rightarrow &|+\rangle
_{A}\prod\limits_{j=1}^{n}|i_{j}\rangle ,  \nonumber \\
|-\rangle _{A}\prod\limits_{j=1}^{n}|i_{j}\rangle  &\rightarrow &|-\rangle
_{A}\prod\limits_{j=1}^{n}e^{i2\theta _{j}\left\langle -\right\vert \left.
i_{j}\right\rangle }|i_{j}\rangle ,
\end{eqnarray}%
which implies that when and only when the control qubit $A$ is in the state $%
|-\rangle $, a phase shift $e^{i2\theta _{j}}$ happens to the state $%
|-\rangle $ of the target qubit $j$ but nothing otherwise (Fig.~2). For $%
\theta _{j}=\pi /2,$ the state transformation~(2) corresponds to a
multi-target-qubit phase gate, i.e., if and only if the control qubit $A$ is
in the state $|-\rangle $, a phase flip from the sign $+$ to $-$ occurs to
the state $|-\rangle $ of each target qubit. Such a multiqubit phase gate
has many applications in QIP~[13-16]. In addition, for $\theta _{j}=\pi
/2^{j},$ the state transformation~(2) corresponds to a multi-target-qubit
phase gate, i.e., if and only if the control qubit $A$ is in the state $%
|-\rangle $, a phase shift $\theta _{j}=\pi /2^{j}$ happens to the state $%
|-\rangle $ of each target qubit. It is noted that this multi-target-qubit
gate is equalivent to a multiqubit gate with different control qubits acting
on the same target qubit (Fig.~3), which is a key element in quantum Fourier
transform~[8,12].

In what follows, our goal is propose a simple method for implementing a
generic unconventional geometric (UG) multi-target-qubit gate described by
Eq.~(1), with one qubit (qubit $A$) simultaneously controlling $n$ target
qubits ($1,2,...,n$) distributed in $n$ cavities ($1,2,...,n$). We believe
that this work is also of interest from the following point of view.
Large-scale QIP usually involves a number of qubits. Placing many qubits in
a single cavity may cause some fundamental problems such as introducing the
unwanted qubit-qubit interaction, increasing the cavity decay, and
decreasing the qubit-cavity coupling strength. In this sense, large-scale
QIP may need to place qubits in multiple cavities and thus require
performing various quantum logic operations on qubits distributed in
different cavities. Hence, it is important and imperative to explore how to
realize multiqubit gates performed on qubits that are spatially-separated
and distributed in different cavities.

As shown below, this proposal has the following features and advantages: (i)
The gate operation time is independent of the number of qubits; (ii) The
proposed multi-target-qubit UG phase gate can be implemented using a
single-step operation; (iii) Only two levels are needed for each qubit,
i.e., no auxiliary levels are used for the state coherent manipulation; (iv)
The proposal is quite general and can be applied to accomplish the same task
with a general types of qubits such as atoms, superconducting qubits,
quantum dots, and NV centers. To the best of our knowledge, this proposal is
the first one to demonstrate that a multi-target-qubit UG phase gate
described by (1) can be achieved with one qubit simultaneously controlling $n
$ target qubits distributed in $n$ cavities.

In this work we will also discuss possible experimental implementation of
our proposal and numerically calculate the operational fidelity for a
three-qubit gate, by using a setup of two superconducting transmission line
resonators each hosting a transmon qubit and coupled to a coupler transmon
qubit. Our numerical simulation shows that highly-fidelity implementation of
a three-qubit (i.e., two-target-qubit) UG phase gate by using this proposal
is feasible with present-day circuit QED technique.

\section*{Results}

\textbf{Model and Hamiltonian.} Consider a system consisting of $n$ cavities
each hosting a qubit and coupled to a common qubit $A$ [Fig.~1(a)]. The
coupling and decoupling of each qubit from its cavity can be achieved by
prior adjustment of the qubit level spacings. For instance, the level
spacings of superconducting qubits can be rapidly adjusted by varying
external control parameters (e.g., magnetic flux applied to the
superconducting loop of a superconducting phase, transmon, Xmon or flux
qubit; see, e.g.,~[38-41]); the level spacings of NV centers can be readily
adjusted by changing the external magnetic field applied along the
crystalline axis of each NV center [42,43]; and the level spacings of
atoms/quantum dots can be adjusted by changing the voltage on the electrodes
around each atom/quantum dot [44]. The two levels of coupler qubit $A$ are
denoted as $|g\rangle _{A}$ and $|e\rangle _{A}$ while those of intracavity
qubit $j$ as $|g\rangle _{j}$ and $|e\rangle _{j}$ $(j=1,2,\cdots ,n)$.
Applying a classical pulse to qubit $A$ and a classical pulse to each
intracavity qubit $j$ [Fig.~1(b),(c)]. For identical qubits, we have $\omega
=\omega _{eg_{A}}=\omega _{eg_{j}}$, where $\omega $ is the pulse frequency
and $\omega _{eg_{A}}$ ($\omega _{eg_{j}}$) is the $|g\rangle
\leftrightarrow |e\rangle $ transition frequency of qubit $A$ (qubit $j$).
The system Hamiltonian in the interaction picture reads (in units of $\hbar
=1$)
\begin{eqnarray}
H_{I} &=&\sum\limits_{j=1}^{n}g_{j}(e^{-i\delta _{j}t}a_{j}^{\dagger }\sigma
_{j}^{-}+h.c.)+\sum\limits_{j=1}^{n}g_{A_{j}}(e^{-i\delta
_{A_{j}}t}a_{j}^{\dagger }\sigma _{A}^{-}+h.c.)  \nonumber \\
&+&\sum\limits_{j=1}^{n}\Omega (\sigma _{j}^{+}+\sigma _{j}^{-})+\Omega
(\sigma _{A}^{+}+\sigma _{A}^{-}),
\end{eqnarray}%
where $a_{j}^{\dagger }$ is the photon creation operator for the mode of
cavity $j$, $\sigma _{A}^{+}~(\sigma _{j}^{+})=|e\rangle _{A}\langle g|$~($%
|e\rangle _{j}\langle g|$) and $\sigma _{A}^{-}~(\sigma _{j}^{-})=|g\rangle
_{A}\langle e|$~($|g\rangle _{j}\langle e|$) are the raising and lowering
operators for qubit $A$ (qubit $j$), $\delta _{j}=\omega _{eg_{j}}-\omega
_{c_{j}}$ and $\delta _{A_{j}}=\omega _{eg_{A}}-\omega _{c_{j}}$ are
detunings (with $\omega _{c_{j}}$ being the frequency of cavity $j$), $%
\Omega $ is the Rabi frequency of the pulse applied to each qubit, $%
g_{A_{j}} $ ($g_{j}$) is the coupling constant of qubit $A$ ($j$) with
cavity $j$. We choose $|\pm \rangle _{j}=(|e\rangle _{j}\pm |g\rangle _{j})/%
\sqrt{2}$ and $|\pm \rangle _{A}=(|e\rangle _{A}\pm |g\rangle _{A})/\sqrt{2}$
as the rotated basis states of qubit $j$ and qubit $A$, respectively.

In a rotated basis \{$|+\rangle _{l},|-\rangle _{l}$\}, one has $\sigma
_{l}^{+}=\left( \widetilde{\sigma }_{z_{l}}-\widetilde{\sigma }_{l}^{+}+%
\widetilde{\sigma }_{l}^{-}\right) /2$ and $\sigma _{l}^{-}=\left(
\widetilde{\sigma }_{z_{l}}+\widetilde{\sigma }_{l}^{+}-\widetilde{\sigma }%
_{l}^{-}\right) /2$, where $\widetilde{\sigma }_{z_{l}}=|+\rangle
_{l}\langle +|-|-\rangle _{l}\langle -|$, $\widetilde{\sigma }%
_{l}^{+}=|+\rangle _{l}\langle -|$, and $\widetilde{\sigma }%
_{l}^{-}=|-\rangle _{l}\langle +|$. Here, $l=1,2,3,\cdots n$, $A$. Hence,
the Hamiltonian~(3) can be expressed as
\begin{eqnarray}
H_{I} &=&\sum\limits_{j=1}^{n}\frac{1}{2}g_{j}[e^{-i\delta
_{j}t}a_{j}^{\dagger }(\widetilde{\sigma }_{z_{j}}+\widetilde{\sigma }%
_{j}^{+}-\widetilde{\sigma }_{j}^{-})+h.c.]  \nonumber \\
&+&\sum\limits_{j=1}^{n}\frac{1}{2}g_{A_{j}}[e^{-i\delta
_{A_{j}}t}a_{j}^{\dagger }(\widetilde{\sigma }_{z_{A}}+\widetilde{\sigma }%
_{A}^{+}-\widetilde{\sigma }_{A}^{-})+h.c.]  \nonumber \\
&+&\sum\limits_{j=1}^{n}\Omega \widetilde{\sigma }_{z_{j}}+\Omega \widetilde{%
\sigma }_{z_{A}}.
\end{eqnarray}%
In a new interaction picture under the Hamiltonian $H_{0}^{\prime
}=\sum\limits_{j=1}^{n}\Omega \widetilde{\sigma }_{z_{j}}+\Omega \widetilde{%
\sigma }_{z_{A}}$, one obtains from Eq.~(4)
\begin{eqnarray}
H_{I}^{\prime} &=&\sum\limits_{j=1}^{n}\frac{1}{2}g_{j}[e^{-i\delta
_{j}t}a_{j}^{\dagger }(\widetilde{\sigma }_{z_{j}}+e^{2i\Omega t}\widetilde{%
\sigma }_{j}^{+}-e^{-2i\Omega t}\widetilde{\sigma }_{j}^{-})+h.c.]  \nonumber
\\
&+&\sum\limits_{j=1}^{n}\frac{1}{2}g_{A_{j}}[e^{-i\delta
_{A_{j}}t}a_{j}^{\dagger }(\widetilde{\sigma }_{z_{A}}+e^{2i\Omega t}%
\widetilde{\sigma }_{A}^{+}-e^{-2i\Omega t}\widetilde{\sigma }%
_{A}^{-})+h.c.].
\end{eqnarray}%
In the strong driving regime $2\Omega \gg \{g_{j},\left\vert \delta
_{j}\right\vert ,g_{A_{j}},\left\vert \delta _{A_{j}}\right\vert \}$, one
can apply a rotating-wave approximation and eliminate the terms that
oscillate with high frequencies. Thus, the Hamiltonian~(5) becomes
\begin{eqnarray}
H_{I}^{\prime}=\sum\limits_{j=1}^{n}\frac{1}{2}g_{j}\widetilde{\sigma }%
_{z_{j}}(e^{-i\delta _{j}t}a_{j}^{\dagger }+h.c.)+\sum\limits_{j=1}^{n}\frac{%
1}{2}g_{A_{j}}\widetilde{\sigma }_{z_{A}}(e^{-i\delta
_{A_{j}}t}a_{j}^{\dagger }+h.c.).
\end{eqnarray}
For simplicity, we set
\begin{eqnarray}
g_{A_{j}}=g_{j},~\delta _{j}=\delta _{A_{j}}.
\end{eqnarray}
The first term of condition~(7) can be achieved by adjusting the position of
qubit $j$ in cavity $j$, and second term can be met for identical qubits.
Thus, the Hamiltonian~(6) changes to
\begin{eqnarray}
H_{eff}=\sum\limits_{j=1}^{n}H_{eff,_{j}}
\end{eqnarray}
with
\begin{eqnarray}
H_{eff,_{j}}=\frac{1}{2}g_{j}(e^{-i\delta _{j}t}a_{j}^{\dagger }+e^{i\delta
_{j}t}a_{j})(\widetilde{\sigma }_{z_{j}}+\widetilde{\sigma }_{z_{A}}),
\end{eqnarray}
where $H_{eff,_{j}}$ is the effective Hamiltonian of a subsystem, which
consists of qubit $A$, intracavity qubit $j$, and cavity $j$. In the next
section, we first show how to use the Hamiltonian~(9) to construct a
two-qubit UG phase gate with qubit $A$ controlling the target qubit $j$. We
then discuss how to use the effective Hamiltonian (8) to construct a
multi-qubit UG phase gate with qubit $A$ simultaneously controlling $n$
target qubits distributed in $n$ cavities.

\textbf{Implementing multiqubit UG phase gates.} Consider a system
consisting of the coupler qubit $A$ and an intracavity qubit $j$, for which $%
|\pm \rangle _{j}$ ($|\pm \rangle _{A}$) are eigenstates of the operator $%
\widetilde{\sigma }_{z_{j}}$ $(\widetilde{\sigma }_{z_{A}})$ with
eigenvalues $\pm 1$. In the rotated basis $\{|+\rangle _{A}|+\rangle
_{j},|+\rangle _{A}|-\rangle _{j},|-\rangle _{A}|+\rangle _{j},|-\rangle
_{A}|-\rangle _{j}\}$, the Hamiltonian~(9) can be rewritten as
\begin{eqnarray}
H_{eff,_{j}}=g_{j}(e^{-i\delta _{j}t}a_{j}^{\dagger }+e^{i\delta
_{j}t}a_{j})\times \left( |+\rangle _{A}|+\rangle _{j}\left\langle
+\right\vert _{A}\left\langle +\right\vert _{j}-|-\rangle _{A}|-\rangle
_{j}\left\langle -\right\vert _{A}\left\langle -\right\vert _{j}\right) ,
\end{eqnarray}
and thus the time evolution operator $U_{Aj}(t)$ corresponding to the
Hamiltonian $H_{eff,_{j}}$ can be expressed as
\begin{eqnarray}
U_{Aj}(t) &=&U_{++,_{j}}(t)|+\rangle _{A}|+\rangle _{j}\left\langle
+\right\vert _{A}\left\langle +\right\vert _{j}+|+\rangle _{A}|-\rangle
_{j}\left\langle +\right\vert _{A}\left\langle -\right\vert _{j}  \nonumber
\\
&+&|-\rangle _{A}|+\rangle _{j}\left\langle -\right\vert _{A}\left\langle
+\right\vert _{j}+U_{--,_{j}}(t)|-\rangle _{A}|-\rangle _{j}\left\langle
-\right\vert _{A}\left\langle -\right\vert _{j},
\end{eqnarray}%
where $U_{++,_{j}}(t)$ and $U_{--,_{j}}(t)$ are given by
\begin{eqnarray}
U_{pp,_{j}}(t) &=&\hat{T}_{j}\exp (-i\int_{0}^{t}H_{pp,_{j}}(\tau )d\tau )
\nonumber \\
&=&\hat{T}_{j}\exp [-ig_{j}\varepsilon _{pp}\int_{0}^{t}(e^{-i\delta
_{j}\tau }a_{j}^{\dagger }+e^{i\delta _{j}\tau }a_{j})d\tau ]  \nonumber \\
&=&\lim_{N\rightarrow \infty }\prod_{n=1}^{N}\exp [-ig_{j}\varepsilon
_{pp}(e^{-i\delta _{j}\tau _{n}}a_{j}^{\dagger }+e^{i\delta _{j}\tau
_{n}}a_{j})\Delta \tau ]  \nonumber \\
&=&\lim_{N\rightarrow \infty }\prod_{n=1}^{N}D[\triangle \alpha _{pp,j}(\tau
_{n})]  \nonumber \\
&=&D(\int_{c}d\alpha _{pp,j})e^{i\theta _{pp,j}},
\end{eqnarray}%
with
\begin{equation}
H_{pp,_{j}}(t)=\left\langle p\right\vert _{A}\left\langle p\right\vert
_{j}H_{eff,_{j}}|p\rangle _{A}|p\rangle _{j}=g_{j}\varepsilon
_{pp}(e^{-i\delta _{j}t}a_{j}^{\dagger }+e^{i\delta _{j}t}a_{j}),
\end{equation}%
where $pp\in \{++,--\},$ $p\in \{+,-\},$ $\varepsilon _{++}=-\varepsilon
_{--}=1,$ $D$ is the displacement operator (for details, see Methods below),
$\hat{T}_{j}$ is the time ordering operator and $\Delta \tau =t/N$ is the
time interval. From Eq.~(12) and Eq.~(31) below, one obtains $d\alpha
_{pp,j}=-ig_{j}\varepsilon _{pp}e^{-i\delta _{j}\tau }d\tau $ and $\theta
_{pp,j}=Im(\int_{c}\alpha _{pp,j}^{\ast }d\alpha _{pp,j})$. Thus, one has
\begin{eqnarray}
\alpha _{pp,j} &=&\int_{c}d\alpha _{pp,j}=\frac{g_{j}\varepsilon _{pp,j}}{%
\delta _{j}}(e^{-i\delta _{j}t}-1),  \nonumber \\
\theta _{pp,j} &=&-\frac{g_{j}^{2}}{\delta _{j}}\int_{0}^{T_{j}}(1-\cos
\delta _{j}t)dt,
\end{eqnarray}%
where $T_{j}$ is the evolution time required to complete a closed path.

If $t=T_{j}$ is equal to $2m_{j}\pi /|\delta _{j}|$ with a positive integer $%
m_{j}$, we have $\int_{c}\alpha _{pp,j}=0$ according to Eq.~(14), which
shows that when cavity $j$ is initially in the vacuum state, then cavity $j$
returns to its initial vacuum state after the time evolution completing a
closed path. Thus, it follows from Eq.~(12) that we have
\begin{equation}
U_{pp,_{j}}(T_{j})=D(0)e^{i\theta _{pp,j}}=e^{i\theta _{pp,j}}.
\end{equation}%
Here $\theta _{pp,j}$ is the total phase given by Eq.~(14), which is
acquired during the time evolution from $t=0$ to $t=T_{j}$. Note that $%
\theta _{pp,j}$ consists of a geometric phase and a dynamical phase.

It follows from Eqs.~(11) and~(15) that the cyclic evolution is described by
\begin{eqnarray}
U_{Aj}(T_{j}) &=&e^{i\theta _{++,j}}|+\rangle _{A}|+\rangle _{j}\left\langle
+\right\vert _{A}\left\langle +\right\vert _{j}+|+\rangle _{A}|-\rangle
_{j}\left\langle +\right\vert _{A}\left\langle -\right\vert _{j}  \nonumber
\\
&+&|-\rangle _{A}|+\rangle _{j}\left\langle -\right\vert _{A}\left\langle
+\right\vert _{j}+e^{i\theta _{--,j}}|-\rangle _{A}|-\rangle
_{j}\left\langle -\right\vert _{A}\left\langle -\right\vert _{j}.
\end{eqnarray}%
Eq. (14) shows that $\theta _{pp,j}$ is independent of index $pp.$ Thus, we
have $\theta _{++,j}=\theta _{--,j}\equiv \theta _{j}.$ Further, according
to Eq. (14), after an integration for $T_{j}$ $=$ $2m_{j}\pi /|\delta _{j}|$
(set above), we have
\begin{equation}
\theta _{j}=-\frac{g_{j}^{2}}{\delta _{j}}T_{j}=\frac{2m_{j}\pi g_{j}^{2}}{%
\delta _{j}^{2}},
\end{equation}%
which can be adjusted by varying the coupling strength $g_{j}$ and detuning $%
\delta _{j}$. Note that a negative detuning $\delta _{j}<0$ (see Fig. 1) has
applied to the last equality of Eq. (17). The unitary operator~(16)
describes a two-qubit UG phase gate operation. For $\theta _{j}\neq 2n\pi $
with an integer $n$, the phase gate is nontrivial. After returning to the
original interaction picture by performing a unitary transformation $U=\exp
\{-i(\Omega \widetilde{\sigma }_{z_{j}}+\Omega \widetilde{\sigma }%
_{z_{A}})T_{j}\}$, we obtain the following state transformations: $|+\rangle
_{A}|+\rangle _{j}\rightarrow e^{i\theta _{j}}e^{-2i\Omega T_{j}}|+\rangle
_{A}|+\rangle _{j},~|+\rangle _{A}|-\rangle _{j}\rightarrow |+\rangle
_{A}|-\rangle _{j},~|-\rangle _{A}|+\rangle _{j}\rightarrow |-\rangle
_{A}|+\rangle _{j},$ and $|-\rangle _{A}|-\rangle _{j}\rightarrow e^{i\theta
_{j}}e^{2i\Omega T_{j}}\newline
|-\rangle _{A}|-\rangle _{j}$, which can be further written as
\begin{eqnarray}
|+\rangle _{A}|+\rangle _{j} &\rightarrow &e^{i\theta _{j}}|+\rangle
_{A}|+\rangle _{j}  \nonumber \\
|+\rangle _{A}|-\rangle _{j} &\rightarrow &|+\rangle _{A}|-\rangle _{j}
\nonumber \\
|-\rangle _{A}|+\rangle _{j} &\rightarrow &|-\rangle _{A}|+\rangle _{j}
\nonumber \\
|-\rangle _{A}|-\rangle _{j} &\rightarrow &e^{i\theta _{j}}|-\rangle
_{A}|-\rangle _{j},
\end{eqnarray}%
where we have set $\Omega T_{j}=k\pi $ ($k$ is a positive integer). For $%
T_{j}=2m_{j}\pi /|\delta _{j}|$, we have $2\Omega =k|\delta _{j}|/m_{j}$.
The result~(18) shows that a two-qubit UG phase gate was achieved after a
single-step operation described above.

Now we expand the above procedure to a multiqubit case. Consider qubit $A$
and $n$ qubits~(1,2,\newline
$\cdots ,n$) distributed in $n$ cavities [Fig.~1(a)]. From Eqs.~(8) and~(9),
one can see that: (i) each term of $H_{eff}$ acts on a different
intra-cavity qubit but the same coupler qubit $A,$ and (ii) any two terms of
$H_{eff},$ corresponding to different $j,$ commute with each other: $%
[H_{eff,_{j}},H_{eff,_{k}}]=0~(j\neq k=1,2,\cdots n)$. Thus, it is
straightforward to show that the cyclic evolution of the cavity-qubit system
is described by the following unitary operator
\begin{equation}
U(T)=\prod_{j=1}^{n}U_{Aj}(T_{j}),
\end{equation}%
where $U_{Aj}(T_{j})$ is the unitary operator given in Eq.~(16), which
characterizes the cyclic evolution of a two-qubit subsystem (i.e., qubit $A$
and intracavity qubit $j$) in the rotated basis $|+\rangle _{A}|+\rangle
_{j},$ $|+\rangle _{A}|-\rangle _{j},$ $|-\rangle _{A}|+\rangle _{j},$ and $%
|-\rangle _{A}|-\rangle _{j}$.

By changing the detunings $\delta _{j}$ (e.g., via prior design of cavity $j$
with an appropriate frequency), one can have
\begin{equation}
m_{1}/\delta _{1}=m_{2}/\delta _{2}=,\cdots ,=m_{n}/\delta _{n},
\end{equation}%
which leads to $T_{1}=T_{2}=,\cdots ,=T_{n}\equiv T,$ i.e., the evolution
time for each of qubit pairs $(A,1)$, $(A,2)$, $\cdots $, and $(A,n)$ to
complete a cyclic evolution is identical. For the setting here, we have $%
\theta _{j}=-\frac{g_{j}^{2}}{\delta _{j}}T$ resulting from Eq.~(17). Hence,
one can easily find from Eqs.~(18) and~(19) that after a common evolution
time $T$, the $n$ two-qubit UG phase gates characterized by a jointed
unitary operator $U(T)$ of Eq.~(19), which have a common control qubit $A$
but different target qubits ($1,2,...,n$), are simultaneously implemented.
As discussed in the introduction, the $n$ two-qubit UG phase gates here are
equivalent to a multiqubit UG phase gate described by Eq.~(1). Hence, after
the above operation, the proposed multiqubit UG phase gate is realized with
coupler qubit $A$ (control qubit) simultaneously controlling $n$ target
qubits ($1,2,\cdots ,n$) distributed in $n$ cavities.

To see the above more clearly, consider implementing a three-qubit
(two-target-qubit) UG phase gate. For three qubits, there are a total number
of eight computational basis states, denoted by $\{|+\rangle _{A}|+\rangle
_{1}|+\rangle _{2},~|+\rangle _{A}|+\rangle _{1}|-\rangle _{2},~\cdots
,~|-\rangle _{A}|-\rangle _{1}|-\rangle _{2}\}$. According to Eqs.~(18) and
~(19), one can obtain a three-qubit UG phase gate, which is described by
\begin{eqnarray}
&|+\rangle _{A}|+\rangle _{1}|+\rangle _{2}\rightarrow &e^{i(\theta
_{1}+\theta _{2})}|+\rangle _{A}|+\rangle _{1}|+\rangle _{2},~~|-\rangle
_{A}|+\rangle _{1}|+\rangle _{2}\rightarrow |-\rangle _{A}|+\rangle
_{1}|+\rangle _{2},  \nonumber \\
&|+\rangle _{A}|+\rangle _{1}|-\rangle _{2}\rightarrow &e^{i\theta
_{1}}|+\rangle _{A}|+\rangle _{1}|-\rangle _{2},~~~~~~~~~|-\rangle
_{A}|+\rangle _{1}|-\rangle _{2}\rightarrow e^{i\theta _{2}}|-\rangle
_{A}|+\rangle _{1}|-\rangle _{2},  \nonumber \\
&|+\rangle _{A}|-\rangle _{1}|+\rangle _{2}\rightarrow &e^{i\theta
_{2}}|+\rangle _{A}|-\rangle _{1}|+\rangle _{2},~~~~~~~~~|-\rangle
_{A}|-\rangle _{1}|+\rangle _{2}\rightarrow e^{i\theta _{1}}|-\rangle
_{A}|-\rangle _{1}|+\rangle _{2},  \nonumber \\
&|+\rangle _{A}|-\rangle _{1}|-\rangle _{2}\rightarrow &|+\rangle
_{A}|-\rangle _{1}|-\rangle _{2},~~~~~~~~~~\ ~~~~|-\rangle _{A}|-\rangle
_{1}|-\rangle _{2}\rightarrow e^{i(\theta _{1}+\theta _{2})}|-\rangle
_{A}|-\rangle _{1}|-\rangle _{2}.
\end{eqnarray}%
As discussed in the introduction, by applying single-qubit operations, this
three-qubit UG phase gate described by Eq. (21) can be converted into a
three-qubit phase gate illustrated in Fig. 2 or Fig. 3 for $n=2$. In the
next section, as an example, we will give a discussion on the experimental
implementation of this three-qubit UG phase gate for the case of $\theta
_{1}=\theta _{2}=\pi /2$. Based on Eq. (17) and for $T_{1}=T_{2}$ (see
above), one can see that the $\theta _{1}=\theta _{2}$ corresponds to $%
g_{1}^{2}/\delta _{1}=g_{2}^{2}/\delta _{2}$, which can be met by adjusting $%
g_{j}$ (e.g., varying the position of qubit $j$ in cavity $j$) or detuning $%
\delta _{j}$ (e.g., prior adjustment of the frequency of cavity $j$) ($j=1,2$%
).

\textbf{Possible experimental implementation.} Superconducting qubits are
important in QIP due to their ready fabrication, controllability, and
potential scalability [38,45,46]. Circuit QED is analogue of cavity QED with
solid-state devices coupled to a microwave cavity on a chip and is
considered as one of the most promising candidates for QIP [45-49]. In
above, a general type of qubit, for both of the intracavity qubits and the
coupler qubit, is considered. As an example of experimental implementation,
let us now consider each qubit as a superconducting transmon qubit and each
cavity as a one-dimensional transmission line resonator (TLR). We consider a
setup in Fig.~4 for achieving a three-qubit UG phase gate. To be more
realistic, we consider a third higher level $\left\vert f\right\rangle $ of
each transmon qubit during the entire operation because this level $%
\left\vert f\right\rangle $ may be excited due to the $\left\vert
e\right\rangle \leftrightarrow \left\vert f\right\rangle $ transition
induced by the cavity mode(s), which will affect the operation fidelity.
From now on, each qubit is renamed \textquotedblleft qutrit" since the three
levels are considered.

When the intercavity crosstalk coupling and the unwanted $\left\vert
e\right\rangle \leftrightarrow \left\vert f\right\rangle $ transition of
each qutrit are considered, the Hamiltonian~(3) is modified as follows
\begin{equation}
h_I=H_I+\Theta _I,
\end{equation}
where $H_I$ is the needed interaction Hamiltonian in Eq.~(3) for $n=2$,
while $\Theta _I$ is the unwanted interaction Hamiltonian, given by
\begin{eqnarray}
\Theta _I &=&\sum_{j=1}^2\widetilde{g}_j\left( e^{i\widetilde{\delta }
_jt}a_j\sigma _{fe_j}^{+}+h.c.\right) +\sum_{j=1}^2\widetilde{g}_{Aj}\left(
e^{i\widetilde{\delta }_{Aj}t}a_j\sigma _{fe_A}^{+}+h.c.\right)+g_{12}\left(
e^{i\Delta t}a_1a_2^{+}+h.c.\right)  \nonumber \\
&+&\sum\limits_{j=1}^2\widetilde{\Omega} ~[e^{i(\omega_{fe_j}-\omega)t}%
\sigma _{fe_j}^{+}+h.c.]+\widetilde{\Omega} ~[e^{i(\omega_{fe_A}-\omega)t}%
\sigma _{fe_A}^{+}+h.c.],
\end{eqnarray}
where $\sigma _{fe_j}^{+}=\left| f\right\rangle _j\left\langle e\right| $
and $\sigma _{fe_A}^{+}=\left| f\right\rangle _A\left\langle e\right| .$ The
first term describes the unwanted off-resonant coupling between cavity $j$
and the $\left| e\right\rangle \leftrightarrow \left| f\right\rangle $
transition of qutrit $j$, with coupling constant $\widetilde{g}_j$ and
detuning $\widetilde{\delta }_j=\omega _{fe_j}-\omega _{c_j}$ [Fig.~5(a,b)],
while the second term is the unwanted off-resonant coupling between cavity $%
j $ and the $\left| e\right\rangle \leftrightarrow \left| f\right\rangle $
transition of qutrit $A$, with coupling constant $\widetilde{g}_{Aj}$ and
detuning $\widetilde{\delta }_{Aj}=\omega _{fe_A}-\omega _{c_j}$
[Fig.~5(c)]. The third term of Eq.~(23) describes the intercavity crosstalk
between the two cavities, where $\Delta =\omega _{c_2}-\omega_{c_1}=\delta
_1-\delta _2$ is the detuning between the two-cavity frequencies and $g_{12}$
is the intercavity coupling strength between the two cavities. The last two
terms of Eq.~(23) describe unwanted off-resonant couplings between the pulse
and the $\left| e\right\rangle \leftrightarrow \left|f\right\rangle $
transition of each qutrit, where $\widetilde{\Omega}$ is the pulse Rabi
frequency. Note that the Hamiltonian~(23) does not involves $\left|
g\right\rangle \leftrightarrow \left| f\right\rangle $ transition of each
qutrit, since this transition is negligible because of $\omega
_{c_j},~\omega\ll \omega _{fg_j},\omega _{fg_A}$ ($j=1,2$) (Fig.~5).

When the dissipation and dephasing are included, the dynamics of the lossy
system is determined by the following master equation
\begin{eqnarray}
\frac{d\rho }{dt} &=&-i\left[ h_{I},\rho \right] +\sum_{j=1}^{2}\kappa _{j}%
\mathcal{L}\left[ a_{j}\right]  \nonumber \\
&+&\sum_{l=1,2,A}\left\{ \Gamma _{l}\mathcal{L}\left[ \sigma _{l}^{-}\right]
+\Gamma _{fe_l}\mathcal{L}\left[ \sigma _{fe_l}^{-}\right]+\Gamma _{fg_l}%
\mathcal{L}\left[ \sigma _{fg_l}^{-}\right] \right\}  \nonumber \\
&+&\sum_{l=1,2,A}\left\{ \Gamma _{l,\varphi f}\left( \sigma _{ff_l}\rho
\sigma _{ff_l}-\sigma _{ff_l}\rho /2-\rho \sigma _{ff_l}/2\right) \right\}
\nonumber \\
&+&\sum_{l=1,2,A}\left\{ \Gamma _{l,\varphi e}\left( \sigma _{ee_l}\rho
\sigma _{ee_l}-\sigma _{ee_l}\rho /2-\rho \sigma _{ee_l}/2\right) \right\}
,\ \ \ \
\end{eqnarray}
where $\sigma _{fg_l}^{-}=\left\vert g\right\rangle _{l}\left\langle
f\right\vert, \sigma _{ee_l}=\left\vert e\right\rangle _{l}\left\langle
e\right\vert ,\sigma _{ff_l}=\left\vert f\right\rangle _{l}\left\langle
f\right\vert ;$ and $\mathcal{L}\left[ \Lambda \right] =\Lambda \rho \Lambda
^{+}-\Lambda ^{+}\Lambda \rho /2-\rho \Lambda ^{+}\Lambda /2,$ with $\Lambda
=a_{j},\sigma _{l}^{-},\sigma _{fe_l}^{-},\sigma _{fg_l}^{-}.$ Here, $\kappa
_{j}$ is the photon decay rate of cavity $a_{j}$ ($j=1,2$). In addition, $%
\Gamma _{l}$ is the energy relaxation rate of the level $\left\vert
e\right\rangle $ of qutrit $l$, $\Gamma _{fe_l}$ ($\Gamma _{fg_l}$) is the
energy relaxation rate of the level $\left\vert f\right\rangle $ of qutrit $%
l $ for the decay path $\left\vert f\right\rangle \rightarrow \left\vert
e\right\rangle (\left\vert g\right\rangle ) $, and $\Gamma _{l,\varphi e}$ ($%
\Gamma _{l,\varphi f}$) is the dephasing rate of the level $\left\vert
e\right\rangle $ ($\left\vert f\right\rangle $) of qutrit $l$ ($l=1,2,A$).

The fidelity of the operation is given by
\begin{equation}
\mathcal{F}=\sqrt{\left\langle \psi _{id}\right\vert {\rho }\left\vert \psi
_{id}\right\rangle },
\end{equation}%
where $\left\vert \psi _{id}\right\rangle $ is the output state of an ideal
system (i.e., without dissipation, dephasing, and crosstalk considered),
while $\rho $ is the final density operator of the system when the operation
is performed in a realistic physical system. As an example, we consider that
qutrit $l$ is initially in a superposition state $1/\sqrt{2}~(|+\rangle
_{l}+|-\rangle _{l})$~$(l=1,2,A)$ and cavity 1 (2) is initially in the
vacuum state. In this case, we have $\left\vert \psi _{id}\right\rangle
=\left\vert \varphi _{id}\right\rangle \otimes |0\rangle _{c1}|0\rangle _{c2}
$, where
\begin{eqnarray}
\left\vert \varphi _{id}\right\rangle  &=&(1/\sqrt{8})(-|+\rangle
_{A}|+\rangle _{1}|+\rangle _{2}+i|+\rangle _{A}|+\rangle _{1}|-\rangle
_{2}+i|+\rangle _{A}|-\rangle _{1}|+\rangle _{2}+|+\rangle _{A}|-\rangle
_{1}|-\rangle _{2}  \nonumber \\
&&+|-\rangle _{A}|+\rangle _{1}|+\rangle _{2}+i|-\rangle _{A}|+\rangle
_{1}|-\rangle _{2}+i|-\rangle _{A}|-\rangle _{1}|+\rangle _{2}-|-\rangle
_{A}|-\rangle _{1}|-\rangle _{2}),
\end{eqnarray}
which is obtained based on Eq.~(21) and for $\theta _{1}=\theta _{2}=\pi /2$%
.

We now numerically calculate the fidelity of the gate operation. Without
loss of generality, consider identical transmon qutrits and cavities.
Setting $m_{1}=1$ and $m_{2}=2,$ we have $\delta _{2}=2\delta _{1}$ because
of Eq.~(20), which corresponds to $g_{1}/g_{2}=1/\sqrt{2}$ for $\theta
_{1}=\theta _{2}$. In order to satisfy the relation $2\Omega \gg |\delta
_{2}|$ and $2\Omega =k|\delta _{2}|/2$, we set $k=12$. In addition, we have $%
\widetilde{g}_{j}\sim \sqrt{2}g_{j}$, $\widetilde{g}_{A_{j}}\sim \sqrt{2}%
g_{A_{j}}$ $(j=1,2)$, and $\widetilde{\Omega }\sim \sqrt{2}\Omega $ for the
transmon qutrits [50]. For a transmon qutrit, a ratio $5\%$ of the
anharmonicity between the $|g\rangle \leftrightarrow |e\rangle $ transition
frequency and the $|e\rangle \leftrightarrow |f\rangle $ transition
frequency is readily achieved in experiments. Thus, we set $\widetilde{%
\delta }_{j}=\delta _{j}-0.05\omega _{eg_{j}}$ and $\widetilde{\delta }%
_{A_{j}}=\delta _{j}-0.05\omega _{eg_{A}}$ ($j=1,2$). For transmon qutrits,
the typical transition frequency between two neighbor levels is between 4
and 10 GHz [51]. Therefore, we choose $\omega _{eg_{A}}/2\pi ,\omega
_{eg_{j}}/2\pi \sim 6.5$ GHz. Other parameters used in the numerical
calculation are as follows: $\Gamma _{l,\varphi e}^{-1}=\Gamma _{l,\varphi
f}^{-1}=10.0$ $\mu $s, $\Gamma _{l}^{-1}=30.0$ $\mu $s, $\Gamma
_{fe_{l}}^{-1}=11.5$ $\mu $s, $\Gamma _{fg_{l}}^{-1}=45.0$ $\mu $s $%
(l=1,2,A) $, and $\kappa _{j}^{-1}=15.0$ $\mu $s ($j=1,2$).

To test how the inter-cavity crosstalk affects the gate fidelity, we plot
Fig.~6 for $g_{12}=0,0.1g_{1},0.2g_{1},0.3g_{1}$, which shows the fidelity
versus $\delta _{1}/2\pi $. For simplicity, the dissipation and dephasing of
the system are not considered in Fig.~6. As depicted in Fig.~6, the effect
of the inter-cavity coupling is negligible as long as $g_{12}\leq 0.1g_{1}$.

Figure~7 shows the fidelity versus $\delta _{1}/2\pi $, which is plotted by
setting $g_{12}=0.1g_{1}$ and now taking the systematic dissipation and
dephasing into account. From Fig.~7, one can see that for $\delta _{1}/2\pi
=-3.57$ MHz, a high fidelity 97.0\% is achievable for a three-qubit UG phase
gate. For $\delta _{1}/2\pi =-3.57$ MHz, we have $T=T_{1}=T_{2}=0.28$$\mu $%
s, $g_{1}/2\pi =1.79$ MHz, and $g_{2}/2\pi =2.52$ MHz. The values of $g_{1}$
and $g_{2}$ here are readily available in experiments [52].

The condition $g_{12}\leq 0.1g_{1}$ is easy to satisfy with the
cavity-qutrit capacitive coupling shown in Fig.~4. When the cavities are
physically well separated, the inter-cavity crosstalk strength is $%
g_{12}\sim g_{A_{1}}C_{2}/C_{\Sigma },g_{A_{2}}C_{1}/C_{\Sigma }$, where $%
C_{\Sigma }=C_{1}+C_{2}+C_{q}$ ($C_{q}$ is the qutrit's self-capacitance)
[53,54]. For $C_{1},C_{2}\sim $ 1~fF and $C_{\Sigma }\sim $ 100~fF (typical
values in experiments), one has $g_{12}=0.01g_{1}$. Thus, the condition $%
g_{12}\leq 0.1g_{1}$ is readily achievable in experiments.

Energy relaxation time $T_{1}$ and dephasing time $T_{2}$ of the level $%
\left\vert e\right\rangle $ can be made to be on the order of $20-60$ $\mu $%
s for state-of-the-art transom devices~[55]. For transmon qutrits, we have
the energy relaxation time $T_{1}^{^{\prime }}\sim T_{1}/2$ and dephasing
time $T_{2}^{\prime }\sim T_{2}$ of the level $\left\vert f\right\rangle ,$
which are comparable to $T_{1}$ and $T_{2},$ respectively. With $\omega
_{eg_{A}}/2\pi ,\omega _{eg_{j}}/2\pi \sim 6.5$ GHz chosen above, we have $%
\omega _{c1}/2\pi \sim 6.503$ GHz and $\omega _{c2}/2\pi \sim 6.507$ GHz.
For the cavity frequencies here and the values of $\kappa _{1}^{-1}$ and $%
\kappa _{2}^{-1}$ used in the numerical calculation, the required quality
factors for the two cavities are $Q_{1}\sim 6.126\times 10^{5}$ and $%
Q_{2}\sim 6.130\times 10^{5}$. Note that superconducting coplanar waveguide
resonators with a loaded quality factor $Q\sim 10^{6}$ were experimentally
demonstrated~[56,57]. We have numerically simulated a three-qubit circuit
QED system, which shows that the high-fidelity implementation of a
three-qubit UG phase gate is feasible with current circuit QED technique.

\section*{Discussion}

A simple method has been presented to realize a generic unconventional
geometric phase gate of one qubit simultaneously controlling $n$
spatially-separated target qubits in circuit QED. As shown above, the gate
operation time is independent of the number $n$ of qubits. In addition, only
a single step of operation is needed and it is unnecessary to employ
three-level or four-level qubits and not required to eliminate the dynamical
phase, therefore the operation is greatly simplified and the experimental
difficulty is significantly reduced. Our numerical simulation shows that
highly-fidelity implementation of a two-target-qubit unconventional
geometric phase gate by using this proposal is feasible with the present
circuit QED technique. The proposed multiqubit gate is generic, which, for
example, can be coonverted into two types of important multi-target-qubit
phase gates useful in QIP and quantum Fourier transform. This proposal is
quite general and can be applied to accomplish the same task with various
types of qubits such as atoms, quantum dots, superconducting qubits, and NV
centers.

\section*{Methods}

\textbf{Geometric phase.} Geometric phase is induced due to a displacement
operator along an arbitrary path in phase space~[58,59]. The displacement
operator is expressed as
\begin{eqnarray}
D(\alpha )=e^{\alpha a^{\dagger }-\alpha ^{\ast }a},
\end{eqnarray}
where $a^{\dagger }$ and $a$ are the creation and annihilation operators of
an harmonic oscillator, respectively. The displacement operators satisfy
\begin{eqnarray}
D(\alpha _{1})D(\alpha _{2})=D(\alpha _{1}+\alpha _{2})e^{i{Im}(\alpha
_{1}\alpha _{2}^{\ast })}.
\end{eqnarray}
For a path consisting of $N$ short straight sections $\Delta \alpha _{j}$,
the total operator is
\begin{eqnarray}
D_{t} &=&D(\Delta \alpha _{N})\cdots D(\Delta \alpha _{j})  \nonumber \\
&=&D(\sum\limits_{j=1}^{N}\Delta \alpha _{j})\exp [i{Im}(\sum%
\limits_{j=2}^{N}\Delta \alpha _{j}\sum\limits_{k=1}^{j-1}\Delta \alpha
_{k}^{\ast })].
\end{eqnarray}%
An arbitrary path $c$ can be approached in the limit $N\rightarrow \infty .$
Therefore, Eq.~(29) can be rewritten as
\begin{eqnarray}
D_{t}=D(\int_{c}d\alpha )e^{i\Theta }
\end{eqnarray}
with
\begin{eqnarray}
\Theta ={Im}({\int_{c}\alpha ^{\ast }d\alpha }).
\end{eqnarray}
For a closed path, we have
\begin{eqnarray}
D_{t}=D(0)e^{i\Theta }=e^{i\Theta },
\end{eqnarray}
where $\Theta $ is the total phase which consists of a geometric phase and a
dynamical phase~[22]. In above, equations~(27-32) have been adopted for
realizing an UG phase gate of one qubit simultaneously controlling $n$
target qubits.

\begin{addendum}

\item[Acknowledgments]

C. P. Yang was supported in part by the National Natural Science Foundation
of China under Grant Nos. 11074062 and 11374083, the Zhejiang Natural
Science Foundation under Grant No. LZ13A040002, and the funds from Hangzhou
Normal University under Grant Nos. HSQK0081 and PD13002004. This work was
also supported by the funds from Hangzhou City for the Hangzhou-City Quantum
Information and Quantum Optics Innovation Research Team.

\item[Author contributions]
T.L and C.P.Y conceived the idea. X.Z.C carried out all calculations under the guidance of Q.P.S and C.P.Y. All the authors discussed the results. T. L and C. P. Y  contributed to the writing of the manuscript.

\item[Additional information]
Competing financial interests: The authors declare no competing financial
interests.
\end{addendum}

\clearpage

\textbf{Figure 1:} (a) Diagram of a coupler qubit $A$ and $n$ cavities each
hosting a qubit. A blue square represents a cavity while a green dot labels
a qubit placed in each cavity, which can be an atom or a solid-state qubit.
The coupler qubit $A$ can be an atom or a quantum dot, and can also be a
superconducting qubit capacitively or inductively coupled to each cavity.
(b) Cavity $j$ is dispersively coupled to qubit $j$ (placed in cavity $j$)
with coupling constant $g_{j}$ and detuning $\delta _{j}<0$. (c) The coupler
qubit $A$ dispersively interacts with cavity $j$, with coupling constant $%
g_{Aj}$ and detuning $\delta _{Aj}<0$ ($j=1,2,...,n$). Here, $\delta
_{Aj}=\delta _{j}$, which holds for identical qubits $A$ and $j$. \bigskip

\textbf{Figure 2:} (a) Schematic circuit of a phase gate with qubit $A$ (a
black dot) simultaneously controlling $n$ target qubits (squares). (b) This
multiqubit phase gate illustrated in (a) consists of $n$ two-qubit phase
gates, each having a shared control qubit (qubit $A$) but a different target
qubit (qubit $1,2,\cdots ,$or $n$). Here, the element $2\theta _{j}$
represents a phase shift $\exp ({i2\theta }_{j})$, which happens to the
state $|-\rangle $ of target qubit $j$ ($j=1,2,...,n$) when and only when
the control qubit $A$ is in the state $|-\rangle $ but nothing happens
otherwise. For $2\theta _{j}=\pi ,$ this gate corresponds to a
multi-target-qubit phase gate (usedful in QIP [13-16]), i.e., if and only if
the control qubit $A$ is in the state $|-\rangle $, a phase flip from the
sign $+$ to $-$ occurs to the state $|-\rangle $ of each target qubit.

\bigskip \textbf{Figure 3:} Schematic circuit of the $n$ successive
two-qubit phase gates in quantum Fourier transform. Here, each two-qubit
phase gate has a shared target qubit (qubit $A$) but a different control
qubit (qubit $1,2,\cdots ,$or $n$). The element $\pi /2^{j}$ represents a
phase shift $\exp ({i\pi /2^{j}})$, which happens to the state $|-\rangle $
of target qubit $A$ if and only if the control qubit $j$ is in the state $%
|-\rangle $ ($j=1,2,...,n$). For any two-qubit controlled phase gate
described by the transformation $|+\rangle _{A}|+\rangle _{j}\rightarrow
|+\rangle _{A}|+\rangle _{j},$ $|+\rangle _{A}|-\rangle _{j}\rightarrow
|-\rangle _{A}|+\rangle _{j},$ $|-\rangle _{A}|+\rangle _{j}\rightarrow
|-\rangle _{A}|+\rangle _{j},$ and $|-\rangle _{A}|-\rangle _{j}\rightarrow
e^{i\phi }|-\rangle _{A}|-\rangle _{j}$, it is clear that the roles of the
two qubits can be interchanged. Namely, the first qubit can be either the
control qubit or the target qubit, and the same applies to the second qubit.
When the second (first) qubit is a control qubit, while the first (second)
qubit is a target, the phase of the state $\left\vert -\right\rangle $ of
the first (second) qubit is shifted by $e^{i\phi }$ when the second (first)
qubit is in the state $\left\vert -\right\rangle $, while nothing happens
otherwise. Thus, the quantum circuit here is equivalent to the circuit
illustrated in Fig. 2 for $2\theta _{j}=\pi /2^{j}$ ($j=1,2,...,n$).

\textbf{Figure 4:} Setup of two cavities (1,2) connected by a
superconducting transmon qubit $A$. Here, each cavity represents a
one-dimensional coplanar waveguide transmission line resonator, qubit $A$ is
capacitively coupled to cavity $j$ via a capacitance $C_{j}$ ($j=1,2$). The
two green dots indicate the two transmon qubits (1,2) embedded in the two
cavities, respectively. The interaction of qubits (1,2) with their cavities
is illustrated in Figs.~5(a) and~(b), respectively. The interaction of qubit
$A$ with the two cavities is shown in Fig.~5(c). Due to three levels for
each qubit considered in our analysis, each qubit is renamed as a qutrit in
Fig.~5. \bigskip

\textbf{Figure 5:} Schematic diagram of qutrit-cavity interaction. (a)
Cavity 1 is coupled to the $\left\vert g\right\rangle \leftrightarrow
\left\vert e\right\rangle $ transition with coupling strength $g_{1}$ and
detuning $\delta _{1}$, but far-off resonant with the $\left\vert
e\right\rangle \leftrightarrow \left\vert f\right\rangle $ transition of
qutrit 1 with coupling strength $\widetilde{g}_{1}$ and detuning $\widetilde{%
\delta }_{1}$. (b) Cavity $2$ is coupled to the $\left\vert g\right\rangle
\leftrightarrow \left\vert e\right\rangle $ transition with coupling
strength $g_{2}$ and detuning $\delta _{2},$ but far-off resonant with the $%
\left\vert e\right\rangle \leftrightarrow \left\vert f\right\rangle $
transition of qutrit $2$ with coupling strength $\widetilde{g}_{2}$ and
detuning $\widetilde{\delta }_{2}$. (c) Cavity $1$ ($2$) is coupled to the $%
\left\vert g\right\rangle \leftrightarrow \left\vert e\right\rangle $
transition of qutrit $A$ with coupling strength $g_{A_{1}}$ ($g_{A_{2}}$)
and detuning $\delta _{A_{1}}$ ($\delta _{A_{2}}$); but far-off resonant
with the $\left\vert e\right\rangle \leftrightarrow \left\vert
f\right\rangle $ transition of qutrit $A$ with coupling strength $\widetilde{%
g}_{A_{1}}$ ($\widetilde{g}_{A_{2}}$) and detuning $\widetilde{\delta }%
_{A_{1}}$ $(\widetilde{\delta }_{A_{2}})$. Here, $\delta _{j}=\omega
_{eg_{j}}-\omega _{c_{j}},\widetilde{\delta }_{j}=\omega _{fe_{j}}-\omega
_{c_{j}},\delta _{A_{j}}=\omega _{eg_{A}}-\omega _{c_{j}},$ and $\widetilde{%
\delta }_{A_{j}}=\omega _{fe_{A}}-\omega _{c_{j}}$ ($j=1,2$), where $\omega
_{eg_{j}}$ ($\omega _{fe_{j}}$) is the $\left\vert g\right\rangle
\leftrightarrow \left\vert e\right\rangle $ ($\left\vert e\right\rangle
\leftrightarrow \left\vert f\right\rangle $) transition frequency of qutrit $%
j$, $\omega _{eg_{A}}$ ($\omega _{fe_{A}}$) is the $\left\vert
g\right\rangle \leftrightarrow \left\vert e\right\rangle $ ($\left\vert
e\right\rangle \leftrightarrow \left\vert f\right\rangle $) transition
frequency of qutrit $A$, and $\omega _{c_{j}}$ is the frequency of cavity $j$%
. \bigskip

\textbf{Figure 6:} Fidelity versus $\delta _{1}/2\pi ,$ plotted for
different intercavity coupling strengths but without considering the
systematic dissipation and dephasing for simplicity. \bigskip

\textbf{Figure 7:} Fidelity versus $\delta _{1}/2\pi ,$ plotted for $%
g_{12}=0 $ and $g_{12}=0.1g_{1}$ and by taking the systematic dissipation
and dephasing into account. The parameters used in the numerical simulation
for Figs.~6 and~7 are referred to the text. \bigskip

\clearpage
\begin{figure}[tbp]
\begin{center}
\epsfig{file=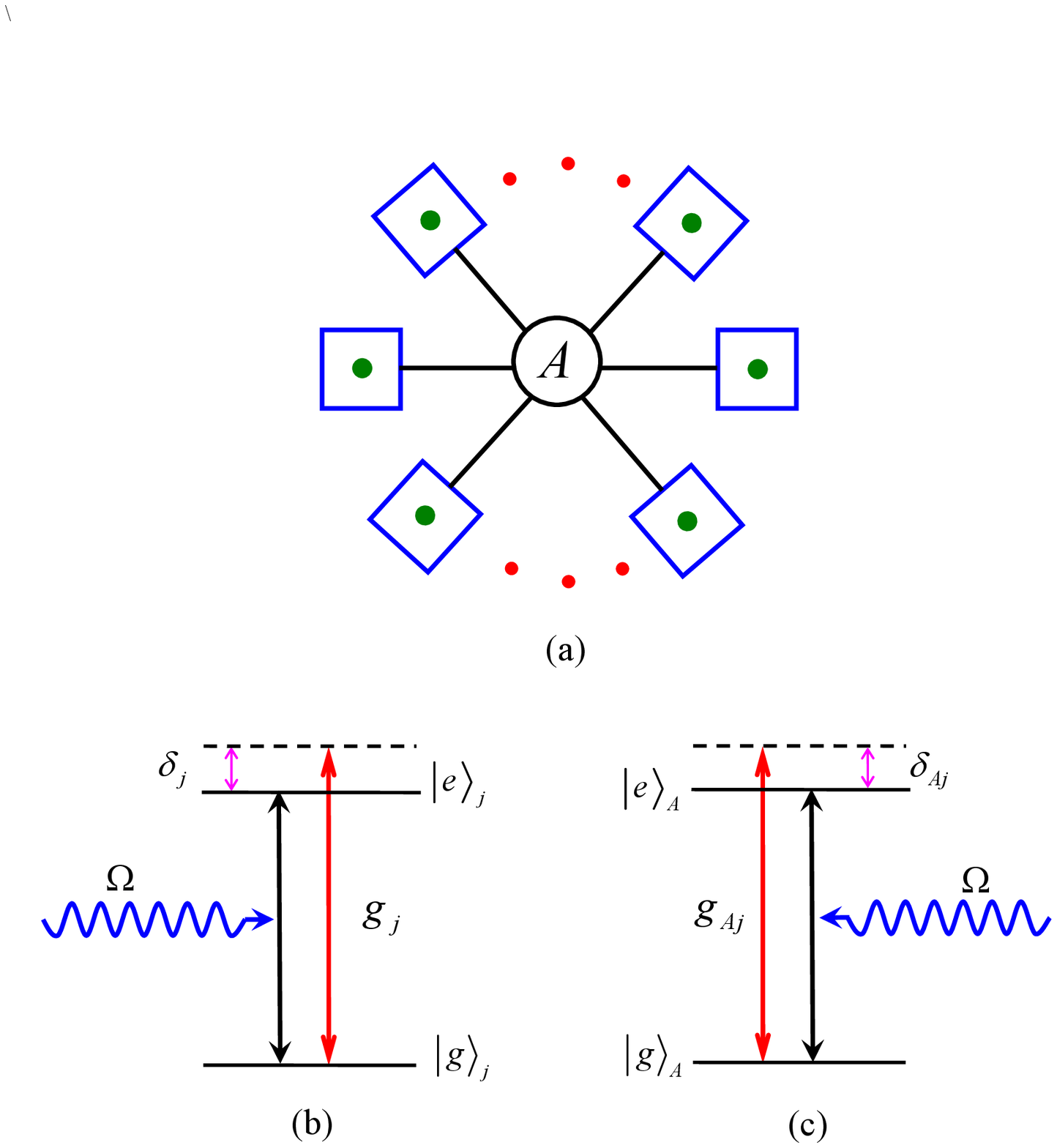,width=13cm}
\end{center}
\caption{}
\label{fig:1}
\end{figure}

\begin{figure}[tbp]
\begin{center}
\epsfig{file=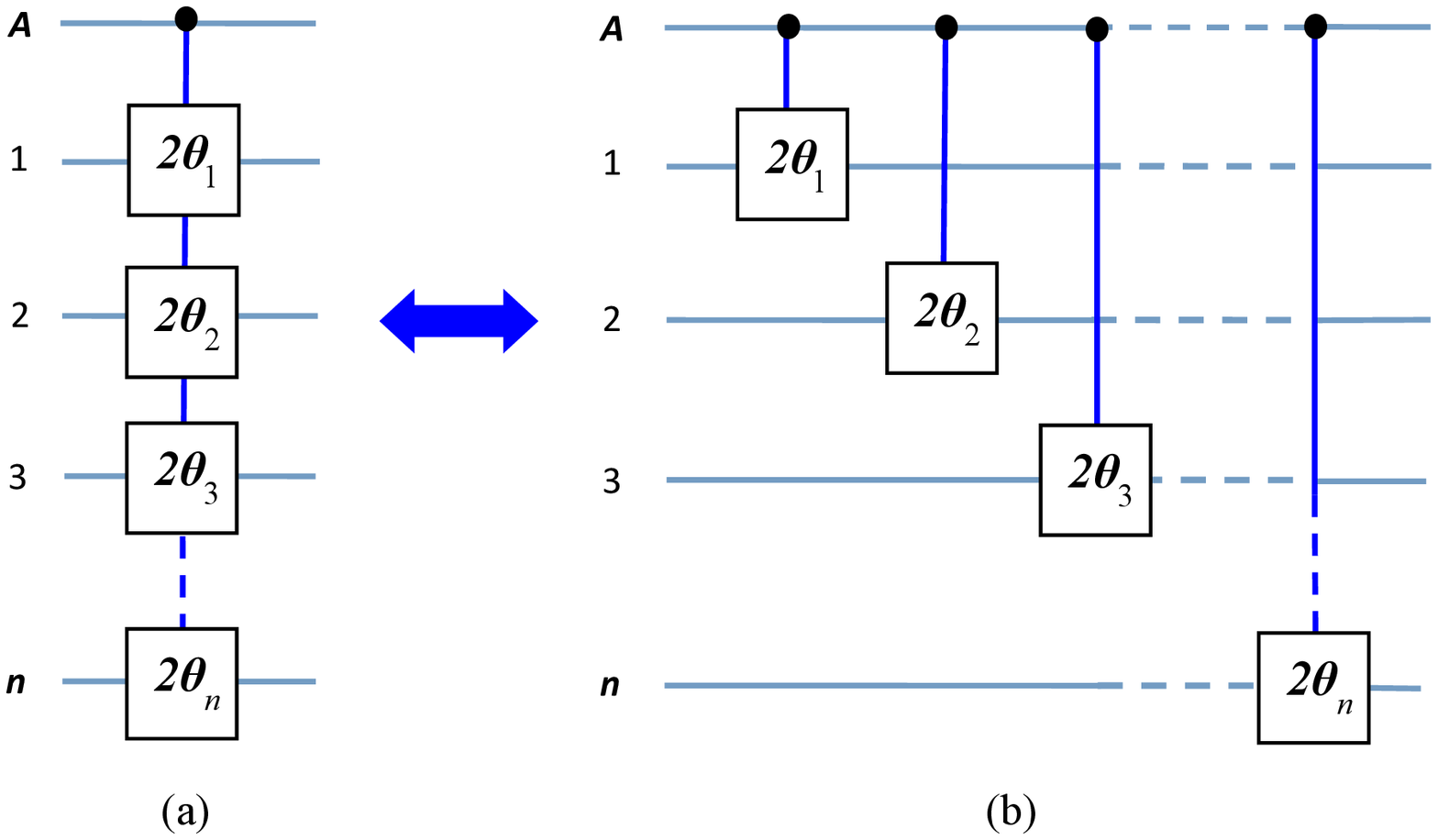,width=13cm}
\end{center}
\caption{}
\label{fig:2}
\end{figure}

\begin{figure}[tbp]
\begin{center}
\epsfig{file=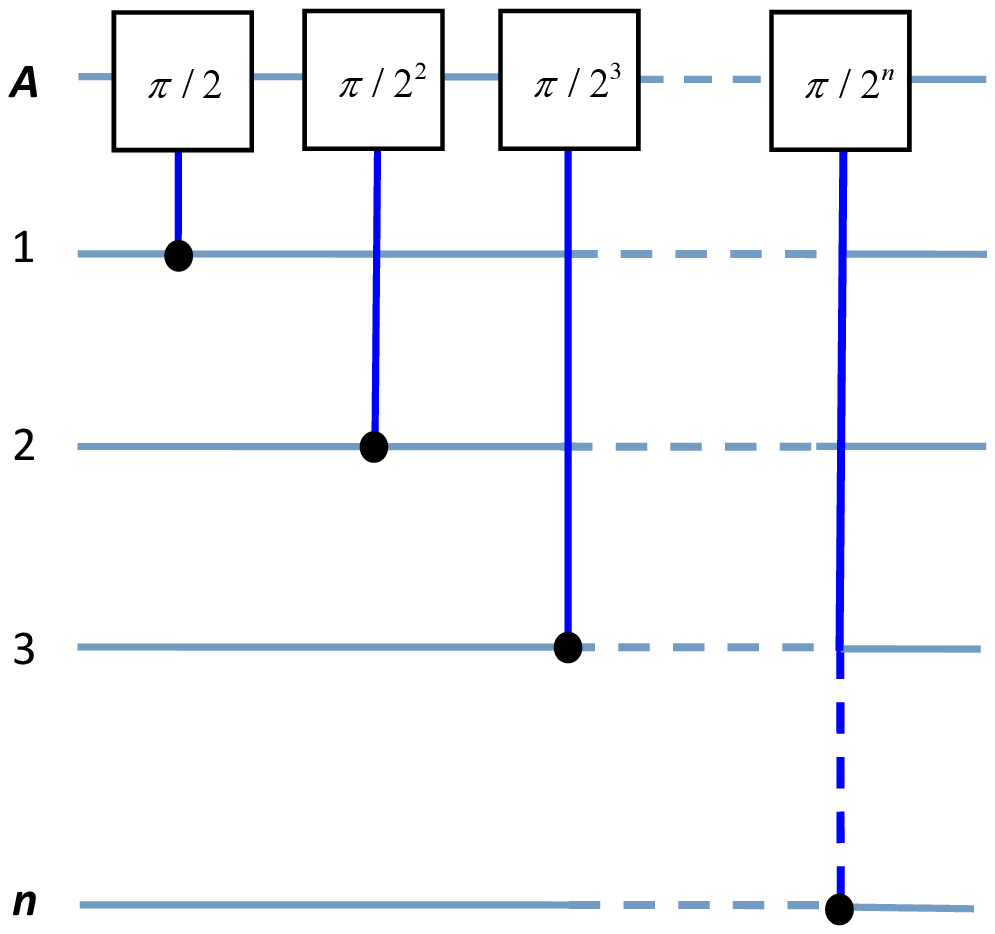,width=15.5cm}
\end{center}
\caption{}
\label{fig:3}
\end{figure}

\begin{figure}[tbp]
\begin{center}
\epsfig{file=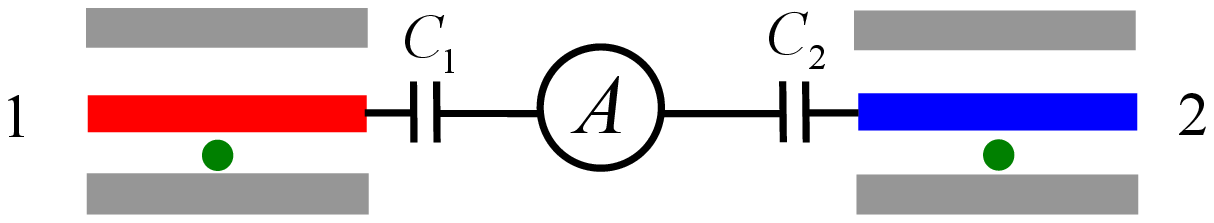,width=15.5cm}
\end{center}
\caption{}
\label{fig:4}
\end{figure}

\begin{figure}[tbp]
\begin{center}
\epsfig{file=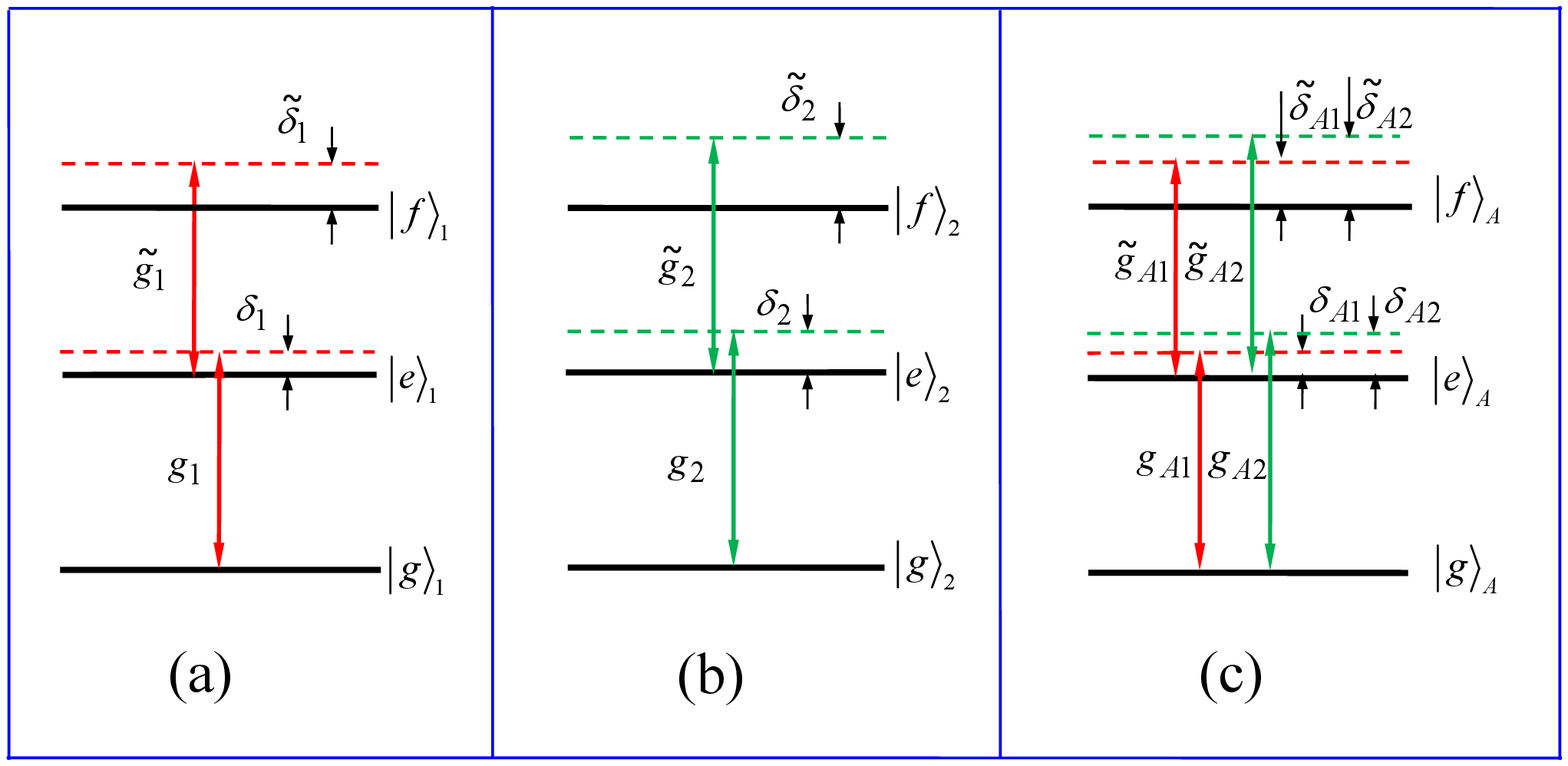,width=14cm}
\end{center}
\caption{}
\label{fig:5}
\end{figure}

\begin{figure}[tbp]
\begin{center}
\epsfig{file=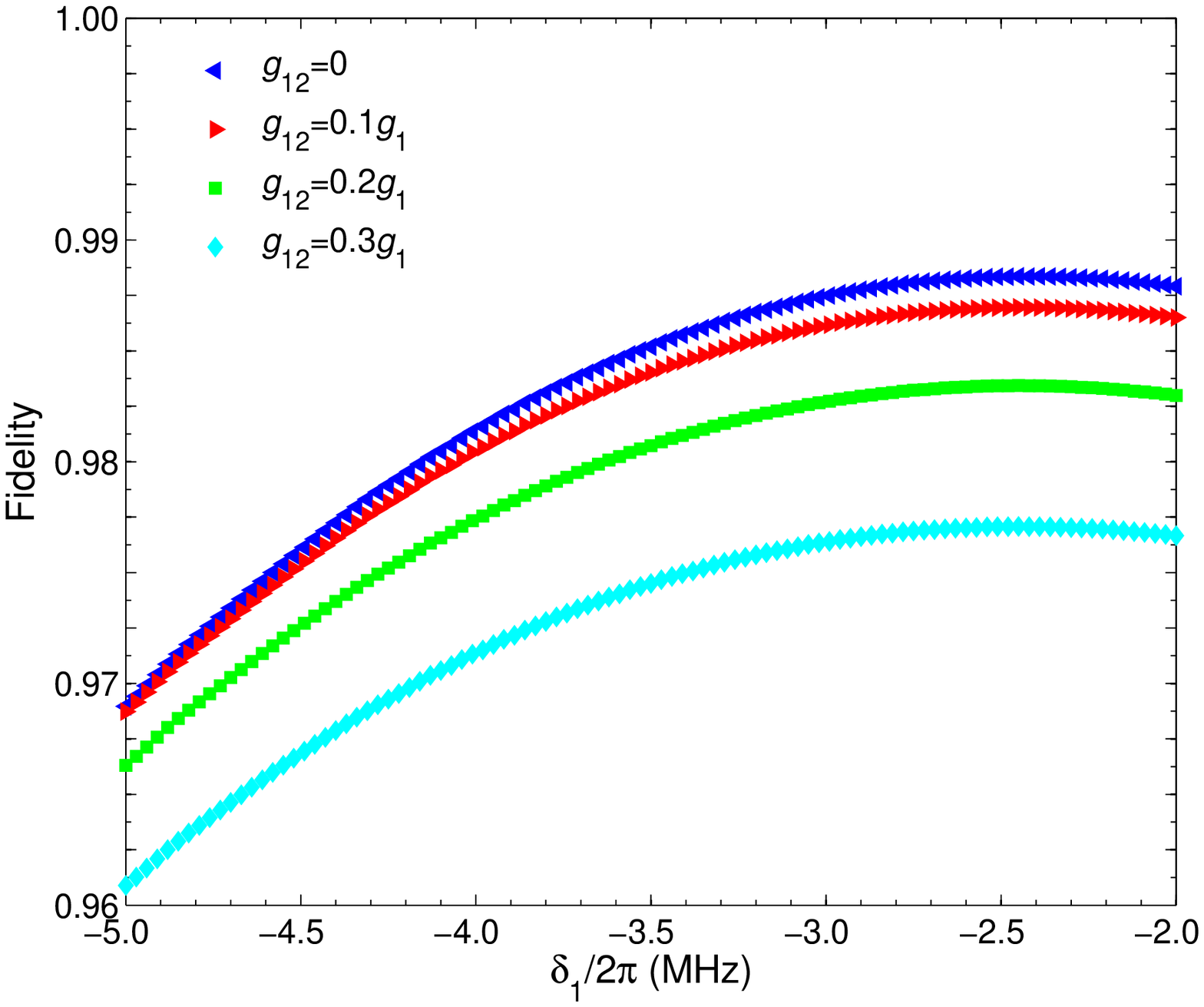,width=13cm}
\end{center}
\caption{}
\label{fig:6}
\end{figure}

\begin{figure}[tbp]
\begin{center}
\epsfig{file=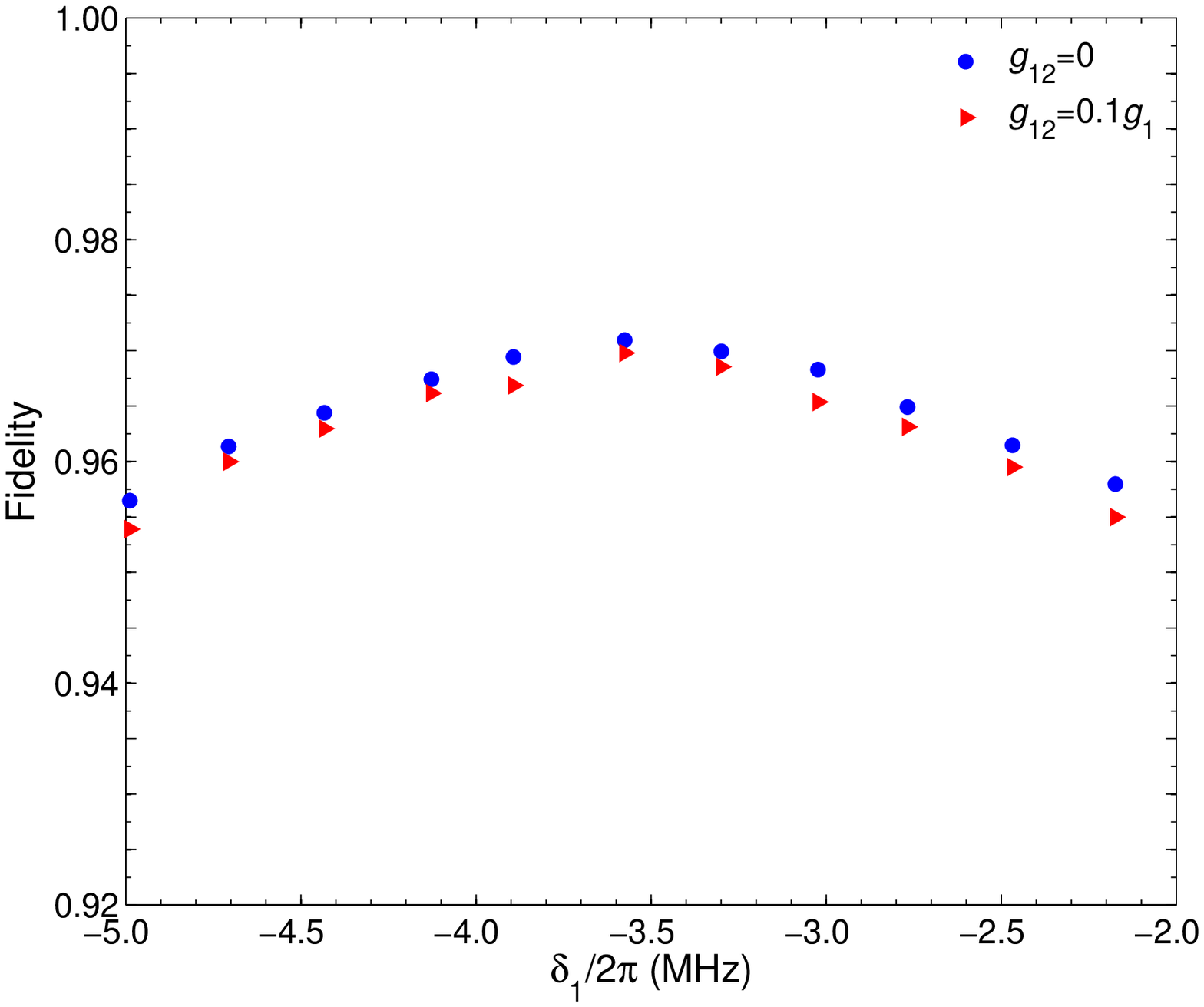,width=13cm}
\end{center}
\caption{}
\label{fig:7}
\end{figure}


\end{document}